\documentclass[11pt]{article}
\usepackage{amsmath}
\usepackage{amssymb}
\usepackage{bbm}
\usepackage{epsfig}
\allowdisplaybreaks[4]
\begin{document}
\title{Charge breaking bounds in the Zee model}
\newcommand{\fslash}[1]{#1 \! \! \! /}

\author{A. Barroso$^{1,}$~\footnote{barroso@cii.fc.ul.pt},
P.M. Ferreira$^{1,}$~\footnote{ferreira@cii.fc.ul.pt} \\
$^1$ Centro de F\'{\i}sica Te\'orica e Computacional, Faculdade de Ci\^encias,\\
Universidade de Lisboa, Av. Prof. Gama Pinto, 2, 1649-003 Lisboa, Portugal}
\date{June, 2005} 
\maketitle
\noindent
{\bf Abstract.} We study the possibility that charge breaking minima occur in 
the Zee model. We reach very different conclusions from those attained in
simpler, two Higgs doublet models, and the reason for this is traced back to the
existence of cubic terms in the potential. A scan of the Zee model's 
parameter space shows that CB is restricted to a narrow region of values of the
parameters. 
\vspace{1cm}

\section{Introduction}

In models with several scalar doublets there is the possibility that minima 
which break charge and/or colour symmetries occur. That happens when 
scalar fields which are not singlets of charge and/or colour acquire a non-zero 
vacuum expectation value. If those minima are deeper than the standard ones,
which respect the $SU(3)_C\times U(1)_{em}$ gauge symmetries, the theory might
tunnel to them, and find itself in a vacuum that breaks charge and/or colour. 
As such, the combinations of 
parameters of the theory that give rise to those minima can be excluded, and the
predictive power of the theory is increased. These charge and colour breaking 
bounds were first introduced by Fr\'ere {\em et al}~\cite{fre} and have been 
extensively used in supersymmetric models, to impose limits on the masses of the
supersymmetric partners, for instance~\cite{cbss}. Recently~\cite{pot10} the 
same idea was applied to two Higgs doublet models (2HDM) without explicit CP 
violation and a very interesting result was deduced: if a minimum that respects 
electric charge conservation and also the CP symmetry (we will call this minimum
the ``normal" one, from this point onwards) exists, then it is the global 
minimum of the theory. In fact, it was shown that only charge breaking saddle 
points can occur in that case, and the value of the potential in those points is
always superior to its value at the normal minimum. In this paper we will 
consider the simplest generalisation of the 2HDM, in which we add an $SU(2)$ 
singlet to the theory. This is the Zee model~\cite{zee}, which was first 
conceived as a means to explain the very low masses of the neutrinos. The Zee
neutrino mass matrix structure can be shown to arise in a more general context, 
through mixing of dimension five operators~\cite{santa1}. We will show that in 
the Zee model charge breaking (CB) minima {\em can and do} develop in parallel 
with normal ones, and sometimes those CB minima are the deepest ones. Then, 
unlike what happens for the 2HDM, tunneling to CB minima is actually possible 
even if the theory starts up in a normal minimum. We will investigate the values
of the potential's parameters for which CB occurs, and  show that CB is mostly 
circumscribed to a specific region of the parameter space.

\section{The Zee model}

The Zee model~\cite{zee} has the great advantage of providing a natural reason
for the smallness of the neutrino masses. Majorana mass terms for the neutrinos 
are generated as radiative corrections at the one-loop level. In addition, the
model can easily be extended~\cite{smi} to incorporate a sterile neutrino, in 
such a way that this four-neutrino version could also accommodate the LSND data.
Unlike the seesaw mechanism, the Zee model does not require new physics at an 
extremely large scale. But it does require a Higgs sector larger than the SM's,
i.e., it requires two scalar doublets, $\Phi_a$ (a=1, 2) with hypercharge $Y = 
1$ and a charged singlet field $\Phi_0$ with hypercharge $Y=-2$~\footnote{Our 
convention for the electric charge operator is $Q\,=\,T_3\,+\,Y/2$.}. In this 
work we are interested in determining the possible vacua of the theory, i.e.
we want to study the scalar potential of the model. The 2HDM scalar potential is
given by a sum of terms quadratic and 
quartic in the fields $\Phi_1$ and $\Phi_2$. There is an extensive literature 
about this potential (see, for instance,~\cite{2hdm}), so we simply summarise  
some of its features: (a) the most general $SU(2)\times U(1)$ gauge invariant 
potential, which explicitly breaks CP conservation, has 14 independent 
real parameters, that can be reduced to 11 with a judicious choice of 
basis~\cite{davh}; (b) 
requiring that the 2HDM does not explicitly break CP one is left with a 10 
parameter potential (though with a choice of basis we can reduce 
this number to 9 independent parameters); this is the potential studied in 
reference~\cite{pot10}, end it has flavour changing neutral currents (FCNC); (c)
if one wishes to prevent the occurrence of FCNC's - or indeed the possibility of
spontaneous breaking of charge or CP symmetries -, it was shown in 
reference~\cite{vel} that the potential has to be invariant under a $Z_2$ or 
$U(1)$ symmetries; one then obtains two different 7-parameter potentials.  

The Zee model usually considered in the literature is based on the 2HDM 
described in point (c) above. {\em A priori} this seems like a good idea, since 
in this manner we prevent potentially dangerous FCNC's and limit the number of 
free parameters of the theory. The problem is that recent 
analysis~\cite{koi,fram} of experimental results in solar and atmospheric 
neutrinos, namely from the SNO experiment, seems to indicate that the minimal 
Zee model cannot account for the data. It has been suggested that extensions of 
the Zee model with extra scalars might do a better job of describing the 
neutrino results (for instance, see~\cite{fram,zeex}). It has also been 
shown~\cite{zeemg} that a Zee model based on the 2HDM potential described in 
point (b) above could accurately reproduce the experiments, and that is the case
we will study here. An obvious problem with this approach is the fact that we 
have increased the number of free parameters in the theory. So, any tool that 
might be useful in reducing the parameter space of the model will be welcome. 
For instance, the authors of refs.~\cite{kane} derived mass bounds on the Higgs 
boson using the triviality condition and (standard) vacuum stability, both for 
the Zee and two higgs dublet models. If we can show that for certain values of 
the parameters deeper CB minima occur, we will be able to impose limits on those
parameters and improve our understanding of the theory. To achieve this, we must
look at the minimisation conditions for the normal and CB minima. 

The fields $\{\Phi_0\,,\,\Phi_1\,,\,\Phi_2\}$ have ten real component fields, 
which we number as
\begin{equation}
\Phi_1 = \begin{pmatrix} \varphi_1 + i \varphi_2 \\ \varphi_5 + i \varphi_7
\end{pmatrix} \;\; , \;\; \Phi_2 = \begin{pmatrix} \varphi_3 + i \varphi_4 \\
\varphi_6 + i \varphi_8 \end{pmatrix} \;\; , \;\; \Phi_0 = \begin{pmatrix} 
\varphi_9 + i \varphi_{10}\end{pmatrix} \;\; .
\end{equation}
This convention might seem odd, but it is very convenient for writing the mass
matrices of the scalar particles of the theory. We can build five $SU(2)\times
U(1)$ quadratic invariants with these fields, which are:
\begin{align}
x_0\,\equiv\,|\Phi_0|^2 &= \varphi_9^2 + \varphi_{10}^2\nonumber \\
x_1\,\equiv\,|\Phi_1|^2 &= \varphi_1^2 + \varphi_2^2 + \varphi_5^2 + \varphi_7^2
\nonumber \\
x_2\,\equiv\,|\Phi_2|^2 &= \varphi_3^2 + \varphi_4^2 + \varphi_6^2 + \varphi_8^2
\nonumber \\
x_3\,\equiv\,\mbox{Re}(\Phi_1^\dagger\Phi_2) &= \varphi_1 \varphi_3 + \varphi_2
\varphi_4 + \varphi_5\varphi_6 + \varphi_7\varphi_8 \nonumber \\
x_4 \,\equiv\,\mbox{Im}(\Phi_1^\dagger\Phi_2) &= \varphi_1 \varphi_4 -  
\varphi_2 \varphi_3 + \varphi_5 \varphi_8 - \varphi_6 \varphi_7 \;\;.
\label{eq:x}
\end{align}
The construction of the scalar potential is thus very simple, one must simply 
include all terms linear and quadratic in the variables $x_i$. We define the
charge conjugated states as $\Phi_i^c\,=\,\Phi_i^*$. Under this transformation 
$x_4 \rightarrow -x_4$ and all the remaining $x_i$ are invariant, hence a CP 
conserving potential is
\begin{align}
V_M \;\;=& \;\;a_0\, x_0\, + \,a_1\, x_1\, + \,a_2\, x_2\, + \,a_3 x_3 \,+
\nonumber \\
  & \;\;b_{00}\, x_0^2\, +\, b_{11}\, x_1^2\, +\, b_{22}\, x_2^2\, +\, b_{33}\, 
x_3^2\, +\, b_{44}\, x_4^2\, +\nonumber \\ 
& \;\;b_{01}\, x_0 x_1\, +\,b_{02}\, x_0 x_2\, +\,b_{03}\, x_0 x_3\, +\, b_{12}
\, x_1 x_2\, +\, b_{13}\, x_1 x_3\, + \,b_{23}\, x_2 x_3 \;\; .
\label{eq:pot}
\end{align}
The $a_i$ parameters have dimensions of mass squared, the $b_{ij}$ are 
dimensionless and all the parameters are real. For future convenience let us
introduce here a matrix notation similar to the one used in ref.~\cite{pot10}.
Let $X$ be the vector with components $(x_0\,,\,x_1\,,\,x_2\,,\,x_3\,,\,x_4)$
and $A$ and $B$ the vector and symmetric matrix given by
\begin{equation}
A\;=\;\begin{bmatrix} a_0 \\ a_1 \\ a_2 \\a_3 \\ 0 \end{bmatrix} \;\;\; , \;\;\;
B\;=\;\begin{bmatrix} 2 b_{00} & b_{01} & b_{02} & b_{03} & 0 \\ b_{01} & 
2 b_{11} & b_{12} & b_{13} & 0 \\ b_{02} & b_{12} & 2 b_{22} & b_{23} & 0 \\ 
b_{03} & b_{13} & b_{23} & 2 b_{33} & 0 \\ 0 & 0 & 0 & 0 & 2 b_{44}
\end{bmatrix} \;\; .
\label{eq:ab}
\end{equation}
Therefore, the potential $V_M$ is given by 
\begin{equation}
V_M\;=\; A^T\,X \;+\; \frac{1}{2}\,X^T \,B\,X \;\;\; .
\label{eq:vm}
\end{equation}
However, there is also an $SU(2) \times U(1)$ invariant cubic term that gives
an extra contribution to the potential, namely
\begin{align}
V_C \;\;=& \; M\,(\Phi_1^T\,i\sigma_2\,\Phi_2)\,\Phi_0\;\;+\;\;\mbox{h.c.} 
\nonumber \\
 =& \; 2\,M\,\left[\left(\varphi_1\varphi_6 - \varphi_2\varphi_8 - \varphi_3
\varphi_5 + \varphi_4\varphi_7\right)\,\varphi_9\;-\;\left(\varphi_1\varphi_8
+ \varphi_2\varphi_6 - \varphi_3\varphi_7 - \varphi_4\varphi_5\right)\,
\varphi_{10}\right]\;\;\; ,
\end{align}
where $M$ is a real coupling with dimensions of mass. The scalar potential of 
the Zee model is thus $V = V_M \,+\,V_C$. 

The ``normal" minimum, which preserves charge and colour conservation, occurs 
when the fields $\varphi_5$ and $\varphi_6$ have non-zero vevs, $v_1$ and $v_2$
respectively, such that $v_1^2\,+\,v_2^2\,=\,(246\,\mbox{GeV})^2$. At this
minimum, then, the vector $X$ takes the values $X_N \,=\,(0\,,\,v_1^2\,,\,v_2^2
\,,\, v_1 v_2\,,\,0)$, and the value of the potential is given by~\cite{pot10}
\begin{equation}
V_N\;\;=\;\;\frac{1}{2}\,A^T\,X_N \;\;=\;\; \frac{1}{2}\,\left(a_1\,v_1^2\;+\;
a_2\,v_2^2\;+\;a_3\,v_1\,v_2 \right)\;\;<\;\; 0 \;\;\; .
\label{eq:vn}
\end{equation}

The inclusion of the charged field $\Phi_0$ introduces some changes in the 
theory's scalar mass spectrum. If one computes the second derivatives of the Zee
potential with respect to the fields $\varphi_i$ at the normal minimum, we see 
that the neutral sector of this model is identical to that of the 2HDM. The 
reason for this is quite simple: $\Phi_0$ is an electrically charged field and  
so there are no mixing terms with the neutral components of 
$\Phi_1$ and $\Phi_2$. A different thing happens in the charged sector, where 
there is a mixing among the charged fields caused by the cubic terms
$V_C$. The charged higgs mass matrix is now given by (from the left to the 
right, the columns of this matrix correspond to the fields $\varphi_1$, 
$\varphi_2$, $\varphi_3$, $\varphi_4$, $\varphi_9$ and $\varphi_{10}$)
\begin{equation}
M^2_{H^\pm} \;=\; \begin{bmatrix} 2 V^\prime_1 & 0 & V^\prime_3 & V^\prime_4 & 
2 M v_2 & 0 \vspace{0.2cm} \\ 0 & 2 V^\prime_1 & - V^\prime_4 & V^\prime_3 & 0 
 & -\,2 M v_2 \vspace{0.2cm} \\ 
V^\prime_3 & - V^\prime_4 & 2 V^\prime_2 & 0 & -\,2 M v_1 & 0 \vspace{0.2cm}  \\
V^\prime_4 & V^\prime_3 & 0 & 2 V^\prime_2 & 0 & 2 M v_1 \vspace{0.2cm} \\ 
2 M v_2 & 0 & -\,2 M v_1 & 0 & 2 V^\prime_0 & 0 \vspace{0.2cm} \\ 
0 & -\,2 M v_2 & 0 & 2 M v_1 & 0 & 2 V^\prime_0 
\end{bmatrix} \;\;\; ,
\label{eq:mch}
\end{equation}
where we have introduced the notation $V^\prime_i \,=\,\partial V_M/
\partial x_i$. The $V^\prime_i$, of course, are evaluated at the normal minimum
and one has
\begin{equation}
V^\prime_1 \;=\; -\,\frac{v_2}{2\,v_1}\;V^\prime_3 \;\;\; , \;\;\; V^\prime_2 
\;=\; -\,\frac{v_1}{2\,v_2}\;V^\prime_3\;\;\;,\;\;\; V^\prime_4\;=\;0\;\;\;.
\label{eq:vpr}
\end{equation}
The matrix~\eqref{eq:mch} is clearly a generalization of the analogous matrix
obtained in reference~\cite{pot10} for the 2HDM. This matrix has two zero 
eigenvalues (corresponding to the charged Goldstone bosons) and two doubly
degenerate non-zero ones, given by
\begin{equation}
M^2_{H^\pm_{1,2}} \;\;=\;\; \frac{1}{2}\left[V^\prime_0\,+\,V^\prime_1\,+\,
V^\prime_2 \;\pm\;\sqrt{(V^\prime_1\,+\,V^\prime_2\,-\,V^\prime_0)^2\;+\;
4 M^2 (v_1^2\,+\,v_2^2)}\right] \;\;\; .
\end{equation}

\section{Charge breaking minima}

For charge breaking to occur, some charged scalar field has to acquire a 
non-vanishing vev. One might think that the case $\{\varphi_3\,\neq\, 0\,,\,
\varphi_i\,=\,0,\,\forall_{i\neq 3}\}$ (or others, equivalent to this one, where
a single charged field of the doublet has a non-zero vev) would give rise to a CB
minimum. In fact, it is simple to show that such a choice of vevs for any 
$SU(2)$ doublets does not actually break charge conservation~\cite{ndub}. 
However, giving a 
vev to the $\Phi_0$ singlet definitively gives the photon a mass. Let $\varphi_9
\,=\,\gamma\,\neq\,0$ (or $\varphi_{10}\,\neq\,0$, which is equivalent) and all 
the other 
fields zero. In this case only $x_0 = \gamma^2$ is different from zero, and a 
trivial minimisation of the Zee potential shows that $\gamma^2\,=\,- \,a_0\,/ 
2\,b_{00}$. For the potential to be bounded from below we must have $b_{00}>0$ 
and so this minimum only exists if $a_0 <0$. In that case the value of the 
potential is
\begin{equation}
V_{CB_1} \;\; = \;\; - \frac{a_0^2}{4\,b_{00}} \;\;\; .
\end{equation}
Then, charge breaking does not occur if $V_{CB_1}\;>\;V_N$, which implies that 
the theory's parameters must obey the inequality
\begin{equation}
b_{00}\;\; >\;\; -\,\frac{1}{2}\;\frac{a_0^2}{a_1\,v_1^2\;+\;a_2\,v_2^2\;+\;
a_3\,v_1\,v_2} \;\; >\;\;0 \;\;\; .
\label{eq:cb1}
\end{equation}
Given $\{a_0\,,\,a_1\,,\,a_2\,,\,a_3\,,\,v_1\,,\,v_2\}$, we have thus a 
restriction on how small the parameter $b_{00}$ may be. This CB minimum is 
rather trivial, and it seems to impose constraints on only a few of the theory's
16 free parameters. In ref.~\cite{pot10} we considered a CB minimum of the form 
$\varphi_5 = v^\prime_1$, $\varphi_6 = v^\prime_2$ and $\varphi_3 = \alpha$. It 
is easy to see, though, that for the most 
general choice of parameters of the potential this CB minimum is not always 
possible. The minimisation conditions are $\partial V/
\partial \varphi_i \,=\,0$, $\forall_{i=1,\ldots, 10}$ and we wish to find if a 
solution with only $\{\varphi_3\,,\,\varphi_5\,,\,\varphi_6\}\;\neq\;0$ is 
possible. From $\partial V/\partial \varphi_9\,=\,0$ we obtain $2 M 
v^\prime_1 \alpha\,=\,0$, which, for $\alpha \neq 0$, implies either $M=0$ (a 
case of limited interest, as the model almost reduces to the 2HDM) or 
$v^\prime_1\,=\,0$. From $\partial V/\partial \varphi_{5,6}\,=\,0$, with 
$\varphi_5 \,=\,v^\prime_1\,=\,0$ and $\varphi_6\,=\,v^\prime_2\neq 0$, we 
obtain
\begin{equation}
\left\{\begin{array}{l} a_3\,v^\prime_2\;+\;b_{23}\,{v^\prime_2}^3\;=\; 0 
\vspace{0.2cm}\\ 2\,a_2\,v^\prime_2\;+\;4\,b_{22}\,{v^\prime_2}^3\;=\; 0 
\end{array} \right.
\end{equation}
which can only be satisfied for particular combinations of parameters, namely 
$a_3/b_{23}\,=\, a_2/2\,b_{22}$, with $a_2<0$. 

Let us study the general case and take the parameters of the theory as being 
uncorrelated. The conclusions we will reach should therefore encompass all the
special cases, such as the examples we have just discussed. 
A simple investigation of the minimisation conditions quickly tells us that the 
simplest CB choice of vevs, besides the trivial one considered above, is 
$\varphi_1\,=
\,\rho$, $\varphi_3\,=\,\alpha$, $\varphi_5\,=\,v^\prime_1$, $\varphi_6\,=\,
v^\prime_2$ and $\varphi_9\,=\,\gamma$. This gives $\hat{x}_0\,=\,\gamma^2$, 
$\hat{x}_1\,=\,{v^\prime_1}^2\,+\,\rho^2$, $\hat{x}_2\,=\,{v^\prime_2}^2\,+\,
\alpha^2$, $\hat{x}_3\,=\,v^\prime_1 v^\prime_2\,+\,\rho \alpha$ and $\hat{x}_4\,=\,
0$ (we use the hat to distinguish these $x_i$ from their counterparts at the 
normal minimum). In reference~\cite{pot10} a very useful relation between the
values of the 2HDM potential at the normal and CB stationary points was derived,
namely
\begin{align}
V_{CB}\;-\;V_N \;\; &= \;\;\left( \displaystyle{-\frac{V^\prime_3}{4 x_3}}
\right)\;\left[ (v^\prime_1\,v_2 \;-\;v^\prime_2\,v_1)^2\; + \; \alpha^2\,v_1^2
\right] \nonumber \\
 &= \;\; \displaystyle{\frac{M^2_{H^\pm}}{2\,(v_1^2\,+\,v_2^2)}}\;\left[
(v^\prime_1\,v_2 \;-\; v^\prime_2\,v_1)^2\; + \; \alpha^2\,v_1^2\right] \;\;\; ,
\label{eq:difv}
\end{align}
where $\{v^\prime_1\,,\,v^\prime_2\,,\,\alpha\}$ are the CB vevs and $V^\prime_3
\,=\,\partial V/\partial x_3$ is evaluated at the normal stationary point. 
$M^2_{H^\pm}$ is the squared mass of the charged scalars at the normal 
stationary point. The second line in the equation above makes it plain that if
the normal minimum exists then we will necessarily have $V_{CB}\;-\;V_{N}\;>
\;0$, and tunneling to a deeper CB minimum is impossible. Let us now
generalise this result for the present case. 
As a first step we define the vector $V^\prime$, with components $V^\prime_i 
\,=\, \partial V_M/\partial x_i$ evaluated at the normal minimum, i.e., 
\begin{equation}
V^\prime\;\;=\;\; A\;+\;B\,X_N \;\;\;.
\label{eq:vp}
\end{equation}
At the CB stationary point we define the vectors $X_{CB}$, with components $
{X_{CB}}_i\,=\,\hat{x}_i$, and $\hat{V}^\prime\,=\,\hat{V}^\prime_i\,=
\,\left.\frac{}{}\partial V_M/\partial x_i\right|_{CB}$, that is, 
\begin{equation}
\hat{V}^\prime\;\;=\;\; A\;+\;B\,X_{CB}\;\;\;.
\label{eq:vpc}
\end{equation}
It is easy to see, from the CB stationarity conditions, that we have the 
following relations:
\begin{equation}
\hat{V}^\prime_0\;=\; -\,\frac{V_C}{2\,\hat{x}_0}
\;\; , \;\; 
\hat{V}^\prime_1\;=\;-\,\frac{\hat{x}_2}{2\,\hat{x}_3}\,\hat{V}^\prime_3
\;\; , \;\; 
\hat{V}^\prime_2\;=\;-\,\frac{\hat{x}_1}{2\,\hat{x}_3}\,\hat{V}^\prime_3
\;\; , \;\; 
\hat{V}^\prime_3\;=\;\frac{4\,M^2\,\hat{x}_0\,\hat{x}_3}{V_C}
\;\; , \;\;\hat{V}^\prime_4\;=\;0 \;\;\; .
\label{eq:vhat}
\end{equation}
$V_C$ is now evaluated
at the CB stationary point and is equal to $V_C \,=\,2\,M\,\gamma\,(\rho v_2^\prime\,-
\,\alpha v_1^\prime)$. Since the potential is a polynomial with quadratic, cubic
and quartic terms, it is trivial to show that its value at any stationary point
is given by half of the quadratic terms plus a quarter of the cubic ones. Then, 
it follows that 
\begin{equation}
V_{CB}\;\;=\;\; \frac{1}{2}\,X_{CB}^T\,A\;+\;\frac{1}{4}\,V_C \;\;\; .
\label{eq:vcb}
\end{equation}
From eqs.~\eqref{eq:vp} and~\eqref{eq:vpc} we obtain
\begin{equation}
\left\{\begin{array}{l}  X_{CB}^T\,V^\prime\;\;=\;\; X_{CB}^T\,A\;+\;X_{CB}^T\,
B\,X_{N} \vspace{0.2cm}\\ 
X_{N}^T\,\hat{V}^\prime\;\;=\;\; X_{N}^T\,A\;+\;X_{N}^T\,B\,X_{CB}
 \end{array}   \right. \;\;\; .
\label{eq:xx}
\end{equation}
Because the matrix $B$ is symmetric, we have $X_{CB}^T\,B\,X_{N}\;=\;X_{N}^T\,B
\,X_{CB}$. Then, subtracting these equations, and using equations~\eqref{eq:vn} 
and~\eqref{eq:vcb} we obtain
\begin{equation}
X_{CB}^T\,V^\prime\;-\;X_{N}^T\,\hat{V}^\prime\;\;=\;\; X_{CB}^T\,A\;-\;
X_{N}^T\,A\;\;=\;\; 2\,(V_{CB}\;-\;V_N)\;-\; \frac{1}{2}\,V_C \;\;\;.
\label{eq:int}
\end{equation}
On the other hand, from eqs.~\eqref{eq:vpr}, we have
\begin{align}
X_{CB}^T\,V^\prime &= \;\; \gamma^2\,V^\prime_0 \;+\; ({v^\prime_1}^2 \,+\,
\rho^2)\,V^\prime_1 \;+\; ({v^\prime_2}^2 \,+\, \alpha^2)\,V^\prime_2 \;+\; 
(v^\prime_1 v^\prime_2\,+\,\rho \alpha)\,V^\prime_3 \nonumber \\
 &= \;\; \gamma^2\,V^\prime_0 \;-\; \frac{V^\prime_3}{2\,x_3}\,\left[(v^\prime_1
\, v_2\;-\;v^\prime_2\,v_1)^2\; + \; (v_1\,\rho\;-\;v_2\,\alpha)^2 \right]\;\;\; ,
\end{align}
and~\eqref{eq:vhat} give 
\begin{align}
X_{N}^T\,\hat{V}^\prime &= \;\; v_1^2\,\hat{V}^\prime_1 \;+\; v_2^2\,
\hat{V}^\prime_2 \;+\;  v_1 v_2 \,\hat{V}^\prime_3 \nonumber \\
 &= \;\; -\; \frac{\hat{V}^\prime_3}{2\,\hat{x}_3}\,\left[(v^\prime_1\, v_2\;-\;
v^\prime_2\,v_1)^2\; + \; (v_1\,\rho\;-\;v_2\,\alpha)^2 \right]\;\;\; .
\end{align}
Replacing these expressions in~\eqref{eq:int} and using~\eqref{eq:vhat}, we 
finally obtain
\begin{equation}
2\,(V_{CB}\;-\;V_N)\;\;=\;\; \gamma^2\,\left(V^\prime_0 \;-\; \hat{V}^\prime_0
\right)\; + \; \frac{1}{2}\,\left(\frac{\hat{V}^\prime_3}{\hat{x}_3}\;-\;
\frac{V^\prime_3}{x_3}\right)\,\left[(v^\prime_1\, v_2\;-\;v^\prime_2\,v_1)^2\; 
+ \; (v_1\,\rho\;-\;v_2\,\alpha)^2 \right]\;\;\; .
\label{eq:difz}
\end{equation}
There is a striking similarity between this expression and the one found for the
2HDM potential, equation~\eqref{eq:difv}. However, whereas~\eqref{eq:difv}
establishes unequivocally that $V_{CB}\,-\,V_{N}\,>\,0$, for the Zee 
model this is not possible. Depending on
the values of the parameters of the theory, we may well have $V^\prime_0\,>\,
\hat{V}^\prime_0$ and $\hat{V}^\prime_3/\hat{x}_3\,>\,V^\prime_3/x_3$, or 
exactly the opposite. Hence, for certain combinations of parameters, it is {\em 
a priori} possible that a CB minimum deeper than the normal minimum exists. It 
is the existence of the cubic terms in the potential that makes the 
simultaneous existence of normal and CB minima possible. In fact, if $M = 0$ 
eqs.~\eqref{eq:vhat} show that $\hat{V}^\prime_3 \,=\,\hat{V}^\prime_0\,=\,
0$, and that $V^\prime_0\,>\,0$ and $-V^\prime_3/x_3\,>\,0$ are the values of 
the two charged higgs masses at the normal minimum. So we trivially obtain, from
eq.~\eqref{eq:difz}, $V_{CB}\,>\,V_N$. Notice that the presence of the cubic 
term makes the Zee model more similar to supersymmetric theories than to the 
2HDM, and one knows that CB can occur in SUSY theories. 

\section{Numerical results}

In view of the result that we have obtained in the previous section, let us try
to determine what combinations of parameters $\{M\,,\, a_i\,,\,b_{ij}\}$ could 
give rise to CB minima deeper than the normal one. It is impossible to solve 
analytically the CB minimisation conditions. Even numerically, this turns out to
be a difficult task. So, we adopt the following strategy: we choose the values 
of the parameters such that $V$ has a normal minimum. For the same parameters we 
now take $V$ as a function of $\{v^\prime_1\,,\,v^\prime_2\,,\,\rho\,,\,\alpha\,
,\,\gamma\}$. Starting from an initial guess for the values of these CB 
vevs, we numerically sample the value of the potential in its neighborhood and 
continue along the direction for which $V$ decreases. At the end of this
evaluation, we compare the value of $V$ that we have reached with $V_N$. If it
is smaller than $V_N$, then this set of parameters is rejected, because it could
give rise to CB. We tried to overcome the dependence of this method on the 
initial guess by performing multiple runs of the algorithm with different 
initial conditions. 

The potential depends on 16 independent parameters. We randomly generate values
for 14 of these parameters, leaving $a_1$ and $a_2$ out of this procedure. The
value of $\tan\beta\,=\,v_2/v_1$ is also used as input (with $0.5\,<\,\tan\beta
\,<\,100$). We impose some restrictions, though. The $b$ parameters are 
chosen as being of the same order of magnitude, since it would not seem natural 
to do otherwise. Here we were inspired by SUSY models, in which all $b_{ij}$ are 
of similar magnitudes. Likewise, we take the $\{M\,,\,a_i \}$ parameters to be 
of the same order. The intervals of variation for the parameters are: 
$-3\,<\,M\,<3$ TeV, $-1\,<\,a_i\,<\,1$ $(\mbox{TeV})^2$ and $0.01 \,<\,b_{ij}\,
< \,10$. Having chosen these parameters, we use the values of $v_1\,=\,v\,\cos
\beta$ and $v_2\,=\,v\,\sin\beta$ (with $v\,=\,246$ GeV) and the minimisation 
conditions of the normal minimum, eqs.~\eqref{eq:vpr}, to determine the values 
of $a_1$ and $a_2$, given by
\begin{align}
a_1 &=\;\;-\,\frac{1}{2\,v_1}\,\left[a_3\, v_2\,+\, 4\, b_{11}\, v_1^3\,+\, 2\,
\left(b_{12} + b_{33}\right)\,v_2^2\, v_1\,+\,3\, b_{13}\, v_1^2\, v_2\,+\, 
b_{23}\, v_2^3 \right] \nonumber \vspace{0.2cm}\\
a_2 &=\;\;-\,\frac{1}{2\,v_2}\,\left[a_3 \,v_1\,+\, 4\, b_{22}\, v_2^3\,+\, 
2\,\left(b_{12} + b_{33}\right)\,v_1^2\, v_2\,+\,3\, b_{23}\, v_2^2\, v_1\,+\, 
b_{13} \,v_1^3 \right] \;\;\; .
\end{align}
With the values of these parameters we compute the masses of the scalar 
particles, using the expression~\eqref{eq:mch} for the charged 
scalars and the formulae (37) and (38) of reference~\cite{pot10} for the
neutral scalars. We accept only those sets of parameters for which $(80)^2\;<\;
m^2_{H_i}\;<\;(1000)^2$ $(\mbox{GeV})^2$. This interval of variation for the 
squared masses was selected so that they are positive, thus ensuring we are at 
the normal minimum, and larger than the values already ruled out by experimental
searches and inferior to the usual triviality bounds.  

It is possible that, for some combinations of $b_{ij}$, the quartic terms of the
Zee potential become arbitrarily negative when some of the fields tend to 
infinity. This corresponds to a potential which is unbounded from below (UFB)
and this combination of $b_{ij}$ should be rejected on physical grounds. In 
ref.~\cite{pot10} we gave some necessary conditions on the $b$'s to prevent the 
occurrence of UFB directions in the 2HDM, and similar conditions were deduced
and used in this work. The procedure to find them is straightforward: take one
or more of the $\varphi_i$ fields to infinity and determine for which values of 
$b_{ij}$ we might obtain $V\,\rightarrow\, -\,\infty$. The use of 
polar/spherical coordinates is very helpful to study these limits. The most 
trivial conditions one obtains in this fashion are $\{b_{00}\,,\,b_{11}\,,\,
b_{22}\}\,\geq\,0$. The most stringent UFB bounds were found 
taking $\{\varphi_1\,,\,\varphi_3\,,\,\varphi_9\}\, \rightarrow\,\infty$ and 
parameterizing these fields as $\varphi_1\,= \,r\,\sin\alpha\,\cos\theta$, 
$\varphi_3\,= \,r\,\sin\alpha\,\sin\theta$, $\varphi_9\,= \,r\,\cos\alpha$,
with $r\,\rightarrow\,+\infty$ and $\{\alpha\,,\,\theta\}$ arbitrary angles.
Requiring that the limit of the quartic terms of the potential is larger or
equal to zero produces the condition $f(\alpha,\theta)\,\leq\,g(\alpha,\theta)$,
$\forall_{\alpha,\theta}$, with
\begin{align}
f(\alpha,\theta) &= -\,\frac{\sin 2\theta}{2}\,\left[\frac{b_{03}}{4}\,\sin^2 
2\alpha \;+\; \left(b_{13}\,\cos^2\theta\,+\,b_{23}\,\sin^2\theta\right)\right]
\nonumber \\
g(\alpha,\theta) &= \;b_{00}\,\cos^4\alpha\;+\; \left[b_{11}\,\cos^4\theta\,+\,
b_{22}\,\sin^4\theta\,+\,\frac{1}{4}\,\left(b_{12} + b_{33}\right]\,\sin^2 2
\theta\right)\,\sin^4\alpha\;\;+ \nonumber \\
 & \;\;\;\;\;\;\frac{1}{4}\,\left(b_{01}\,\cos^2\theta\,+\,b_{02}\,\sin^2\theta
\right)\,
\sin^2 2\alpha \;\;\; .
\end{align}
We checked, for several values of $\{\alpha\,,\,\theta\}$, if the values of the
b's obeyed the previous inequality. The UFB conditions found in 
ref.~\cite{pot10} for the 2HDM are still valid for the Zee model and have been
used. Furthermore, it is easy to obtain two additional conditions, 
\begin{align}
b_{01} &\geq \,-2\,\sqrt{b_{00}b_{11}} \nonumber \\
b_{02} &\geq \,-2\,\sqrt{b_{00}b_{22}} \;\;\;.
\end{align}
Although these are only necessary conditions, we found that they eliminate most 
of the UFB potentials. Nevertheless, we will see that UFB potentials still 
appear in our analysis, but only for a small subset of parameters. 

Having required these conditions on the $b$ parameters and rejected 
those sets of $\{M\,,\,a_i\,,\,b_{ij}\}$ that did not produce scalar masses 
within the given interval, we built a set containing nearly 103000 different 
combinations of parameters. Although this a small part of the total
parameter space, we hope that it will serve to illustrate the general CB 
properties of the Zee model potential. For each ``point" in this parameter 
space, we computed the value of the potential at the normal minimum, $V_N$, 
numerically obtained the value of $V_{CB}$ and compared both these values.
Despite our effort to enforce the UFB conditions, we have obtained UFB 
potentials for roughly 2\% of the points. 

\begin{figure}[htb]
\epsfysize=8cm
\centerline{\epsfbox{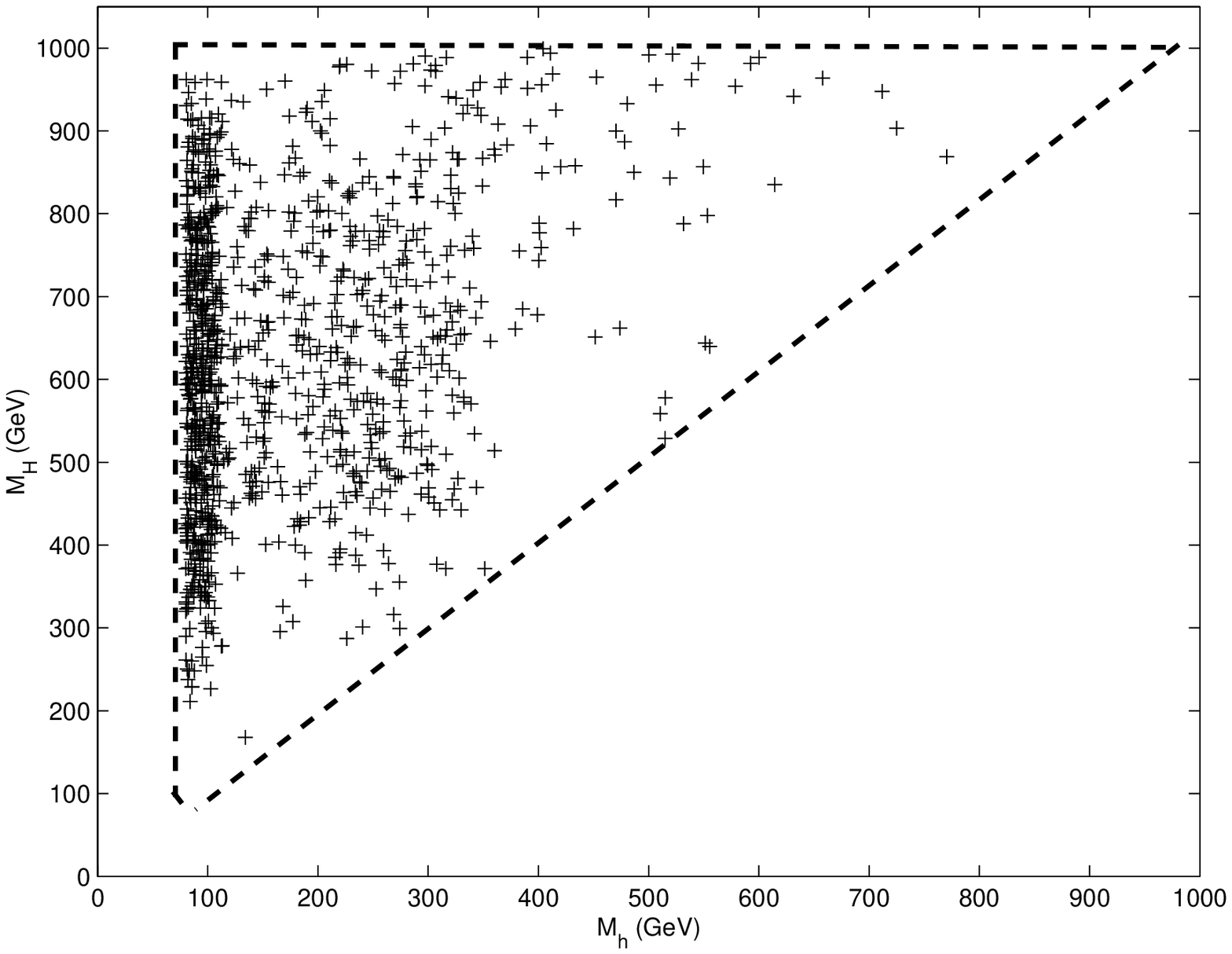}}
\caption{Plot of the mass of the lightest CP-even higgs neutral scalar versus 
the mass of the heaviest; the area delimited by the dashed lines shows the 
whole region of the parameter space. The crosses denote those ``points" for 
which CB occurs. Most of these are 
circunscribed to a narrow interval of low values of $M_h$.}
\label{fig:higgs}
\end{figure}

CB was found to occur, mostly limited to a specific region of parameters. In 
fact, the majority of CB points occurred for large absolute values of the $M$ 
parameter ($|M|>\,500$ GeV), which is easy to understand, given our earlier 
conclusion that the existence of the cubic terms in the potential is 
intimately linked to the possibility of CB. The region where CB was most likely 
to occur has lower values of $\tan\beta$ (inferior to about 10), negative values
of the parameters $b_{33}$, $b_{44}$ and $a_3$ (this last one always superior to
$-(800\;\,\mbox{GeV})^2$) and positive values for $a_1$ and $b_{12}$. Having 
varied the $b$ parameters in an interval spanning three orders of
magnitude (from 0.01 to 10) we found that CB occurs less frequently for the 
larger values of the $b$'s. Although these observations are interesting in 
themselves, looking for the possible impact of CB on physical parameters, such 
as scalar masses, is of much greater interest. We observed that CB occurred for
comparatively smaller values of the lightest neutral (CP even) higgs scalar,
mostly below $\sim$ 350 Gev with the largest concentration occurring for $90\,<
\,M_h\,<\,130$ GeV. This is illustrated in figure~\ref{fig:higgs}, where we plot
the values of $M_h$ versus $M_H$. The area delimited by the dashed lines shows 
the entire zone filled out by our $\sim$ $10^5$ sample of the parameter space. 
Those points for which CB occurs are marked with a cross. It is easy to 
see that CB occurs mostly for a narrow region of values of $M_h$. Similar 
``trends" in the values of the masses are found for the other scalars of the 
theory. CB is concentrated on the relatively small values of the lightest 
charged higgs ($M_{H^\pm_2}\,<\, 300$ GeV), meaning larger values of the 
heaviest charged scalar ($M_{H^\pm_1}\, >\,500$ GeV), and for values of $M_A$, 
the pseudoscalar mass, usually larger than 200 GeV. 

\begin{figure}[htb]
\epsfysize=8cm
\centerline{\epsfbox{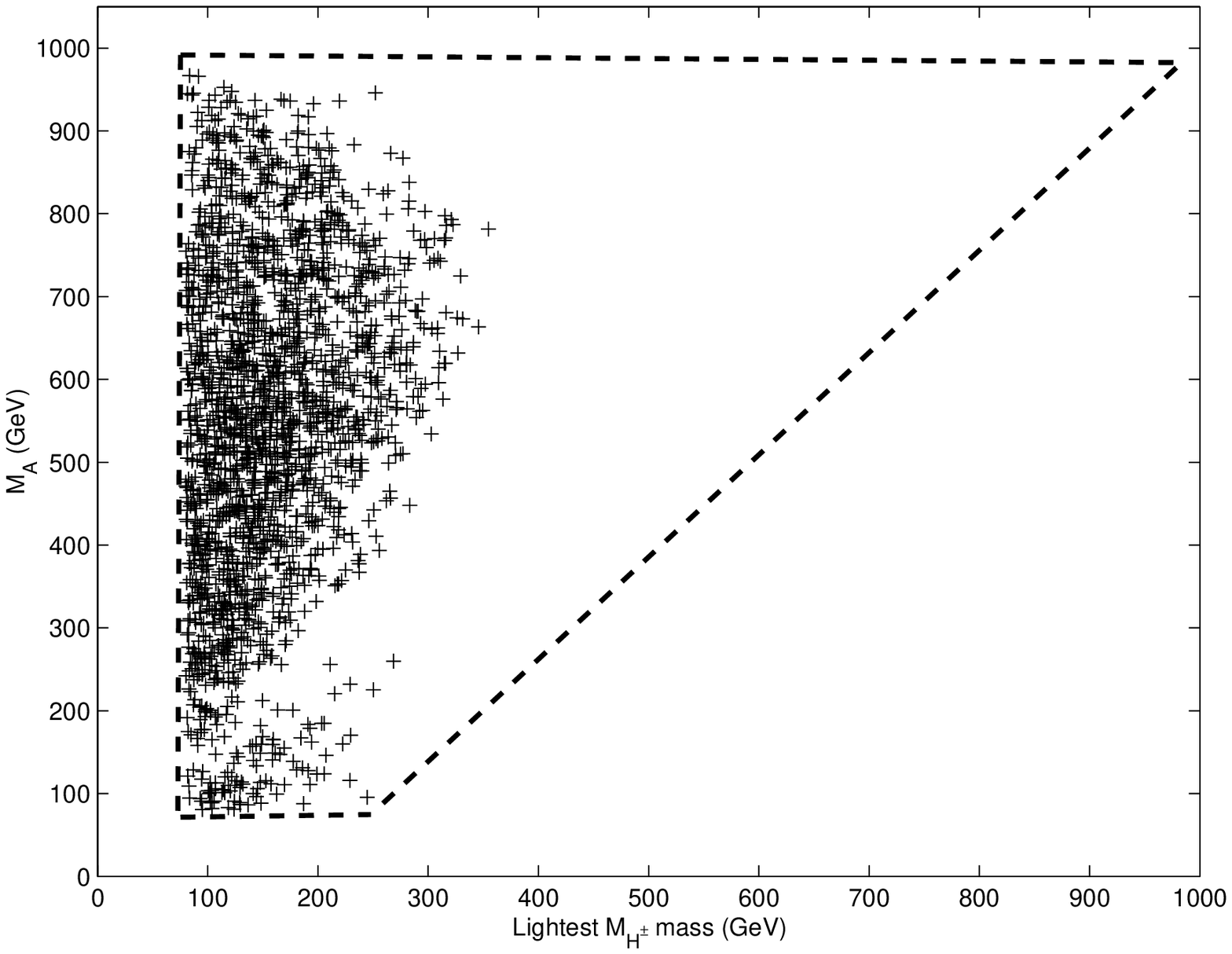}}
\caption{Plot of the mass of the lightest charged higgs scalar versus the mass 
of the pseudoscalar; the area delimited by the dashed lines represents the whole
of the parameter space, and the crosses those points for which CB occurs.} 
\label{fig:ach}
\end{figure}

This initial foray into the CB analysis of the Zee model parameter space
allowed us to identify the areas for which it was more likely to occur. Our 
next step was to generate a second set of parameters (about 53000, this time),
following the same philosophy but with some extra restrictions: namely, $\tan
\beta\leq 10$, $0.01\,<\,b_{ij}\,<\,1$ and imposing $M_h\,<\,500$ GeV. Our 
intent was to narrow the search for CB to those areas where it is more likely
to occur, to gauge its importance in terms of percentage of parameter space
rejected. We found that CB occurred for about 5\% of this second parameter 
space, and confirmed the ``trends" found earlier for the parameters and masses
of the theory. As an illustration of this, the reader can see in 
figure~\ref{fig:ach} how CB occurs for smaller values of the lightest charged
scalar (below 300 GeV, as mentioned above) and mostly for higher values of the
pseudoscalar mass. Again, in this figure the area delimited 
by the dashed lines shows the entire range of allowed values. Unfortunately, 
these plots do {\em not} mean that, after having eliminated all the points of the 
parameter space for which CB occurs, we would be left with ``holes" in this 
space. If it were so, we could make definite restrictions, for instance on the 
higgs' masses. But this is not the case: given the size of the parameter space,
there are {\em many} combinations of $\{M\,,\,a_i\,,\,b_{ij}\}$ which produce 
the same scalar masses. In other words, if in figure~\ref{fig:higgs} we showed 
that CB occurred for $90\,<\,M_h\, <\,130$ GeV, that does not mean that {\em all}
Zee model points which produce masses in that interval were eliminated. On the 
contrary, there are many left which produce such masses. This means that we 
cannot produce a definite statement about the areas of parameter space for which
CB occurs. We can, however, say with a good degree of certainty where CB does 
{\em not} occur: the regions with high values for the masses of the lightest 
neutral and charged higgses ($M_h\,>\,500$ GeV, $M_{H^\pm_2}\,>\,300$ GeV) are 
almost entirely ``clean" of CB.

\section{Conclusions}

We have shown that charge breaking may occur in the Zee model, and
have developed here the algorithm for finding the regions of parameter space
where that may happen. The values of the lightest neutral higgs mass for
which CB is more likely to occur are specially interesting, given that they 
correspond to the values that are currently being probed by our accelerators. 
Our analysis was very general, considering large regions of parameter space with
randomly generated parameters. As was mentioned in the text, the minimal Zee
model does not seem to be compatible with current neutrino data. This work 
focused on more general forms for the two doublets plus one singlet scalar
potential, but extensions of the Zee model, with a larger scalar sector, are
also of interest. The possibility of charge breaking would also arise in those
models, and it would be interesting to perform CB studies on those theories. The
results we obtained here cannot, clearly, be generalised to those other models,
and in fact there is the possibility that, given a particular scalar content, 
the conclusions reached will be very different. For instance, in 
ref.~\cite{santa2} a model with a hyperchargeless triplet majoron was studied,
and it was shown that the vacuum of that theory {\em always} broke charge 
conservation. Extra scalar content of a more ``usual" nature - singlets and 
dublets, charged or not - would make the Zee potential even more similar to the
supersymmetric one, and thus we would expect that charge breaking would still 
occur, as it does in the SUSY case.

Given the results encountered for our study of the non-minimal Zee model, we 
feel that further studies in this area are warranted: specific embodiments of 
the Zee model, with values of parameters specially tuned to explain neutrino 
observational data, should be tested to verify if they are safe from CB. This
is not usually done, since in most applications of the Zee 
model~\cite{koi,fram,zeemg} one does not consider in detail the scalar sector 
of the theory beyond its spectrum of masses and mixing angles. However, such a 
study might help narrow down the range of variation of the theory's parameters 
and, perhaps, shed some light on the generation of neutrino masses. 

\vspace{0.25cm}
{\bf Acknowledgments:} We thank Rui Santos for several interesting discussions.
This work is supported by Funda\c{c}\~ao para a Ci\^encia e Tecnologia under 
contract POCTI/FNU/49523/2002. P.M.F. is supported by FCT under contract 
SFRH/BPD/5575/2001.



\begin{thebibliography}{99}
\bibitem{fre} J.M. Fr\'ere, D.R.T. Jones and S. Raby,
{\em Nucl. Phys.} {\bf B222} (1983) 11.
\bibitem{cbss} L. Alvarez-Gaum\'e, J. Polchinski and M. Wise, {\em Nucl. Phys.}
{\bf B221} (1983) 495;

J.F. Gunion, H.E. Haber and M. Sher, {\em Nucl. Phys.} {\bf B306} (1988) 1;

S. Abel and B. Allanach, {\em Phys. Lett.} {\bf B431} (1998) 339;

OPAL Collaboration, {\em Eur. Phys. Jour} {\bf C7} (1999) 407; {\em ibid.},
{\em Eur. Phys. Jour} {\bf C12} (2000) 567;

M. Antonelli and S. Moretti, {\bf hep-ph/0106332}; 

S.Y. Choi, K. Hagiwara, J.S. Lee, {\em Phys. Lett.} {\bf B529} (2002) 212;

S. Abel, J. Santiago, {\em J. Phys.} {\bf G30} (2004) R83-R111;

A. Djouadi, M. Drees and J.-L. Kneur, {\em Phys. Lett.} {\bf B624} (2005) 60.
\bibitem{pot10} P.M. Ferreira, R. Santos and A. Barroso, {\em Phys. Lett.} {\bf
B603} (2004) 219.
\bibitem{zee} A. Zee, {\em Phys. Lett.} {\bf B93} (1980) 389.
\bibitem{santa1} J.F. Oliver and A. Santamaria, {\em Phys. Rev.} {\bf D65} 
(2002) 033003.
\bibitem{smi} A. Yu. Smirnov and M. Tanimoto, {\em Phys. Rev.} {\bf D55} (1997)
1665.
\bibitem{2hdm}  M. Sher, {\em Phys. Rep.} {\bf 179} (1989) 273;

G.C. Branco, L. Lavoura and J.P. Silva, {\em CP Violation} (Oxford University
Press, Oxford, England, 1999).
\bibitem{davh} S. Davidson and H. Haber, hep-ph/0504050.
\bibitem{vel} J. Velhinho, R. Santos e A. Barroso, {\em Phys. Lett.} {\bf B322}
(1994) 213.
\bibitem{koi} Y. Koide, {\em Phys. Rev.} {\bf D64} (2001) 077301.
\bibitem{fram} P.H. Frampton, M.C. Oh and T. Yoshikawa, {\em Phys. Rev.} {\bf 
D65} (2002) 073014. 
\bibitem{zeex} T. Kitabayashi and M. Yasue, {\em Int. J. Mod. Phys.} {\bf A17}
(2002) 2519;

{\em id.}, {\em Phys. Lett.} {\bf B490} (2000) 236;

D. Chang, W.-Y. Keung and P.B. Pal, {\em Phys. Rev. Lett.} {\bf 61} (1988) 2420;

K.S. Babu, {\em Phys. Lett.} {\bf B203} (1988) 132;

A. Zee, {\em Nucl. Phys.} {\bf B264} (1986) 99.
\bibitem{zeemg}  K. Hasegawa, C.S. Lim and K. Ogure, {\em Phys. Rev.} {\bf D68}
(2003) 053006;

A. Zee, {\em AIP Conf. Proc.} {\bf 689} (2003) 74; hep-ph/0307155. 
\bibitem{kane} S. Kanemura {\em et al}, {\em Phys. Rev.} {\bf D64} (2001) 
053007;

S. Kanemura, T. Kasai and Y. Okada, {\em Phys. Lett.} {\bf B471} (1999) 182.
\bibitem{ndub} A. Barroso, P.M. Ferreira, R. Santos and J.P. Silva, to be 
published.
\bibitem{santa2} A. Santamaria, {\em Phys. Rev.} {\bf D39} (1989) 2715. 
\end{thebibliography}
\end{document}